\documentclass[fleqn,10pt]{wlscirep}
\usepackage[utf8]{inputenc}
\usepackage[T1]{fontenc}
\usepackage{amsmath}
\usepackage{caption}
\usepackage{subcaption}
\usepackage{xcolor}
\title{Using Deep Learning-based Features Extracted from CT scans to Predict Outcomes in COVID-19 Patients}

\author[1,*]{Sai Vidyaranya Nuthalapati}
\author[2]{Marcela Vizcaychipi}
\author[2, 3, 4]{Pallav Shah}
\author[1]{Piotr Chudzik}
\author[1]{Chee Hau Leow}
\author[1]{Paria Yousefi}
\author[1]{Ahmed Selim}
\author[1]{Keiran Tait}
\author[1]{Ben Irving}
\affil[1]{Sensyne Health PLC, Oxford, OX4 4GE, UK}
\affil[2]{Chelsea and Westminster Hospitals, London, SW10 9NH, UK}
\affil[3]{Royal Brompton Hospital, London, SW3 6NP, UK}
\affil[4]{National Heart and Lung Institute, Imperial College,  London, SW3 6LY, UK}
\affil[*]{sai.vidyaranya@sensynehealth.com}

\begin{abstract}
The COVID-19 pandemic has had a considerable impact on day-to-day life. Tackling the disease by providing the necessary resources to the affected is of paramount importance. However, estimation of the required resources is not a trivial task given the number of factors which determine the requirement. This issue can be addressed by predicting the probability that an infected patient requires Intensive Care Unit (ICU) support and the importance of each of the factors that influence it. Moreover, to assist the doctors in determining the patients at high risk of fatality, the probability of death is also calculated. For determining both the patient outcomes (ICU admission and death), a novel methodology is proposed by combining multi-modal features, extracted from Computed Tomography (CT) scans and Electronic Health Record (EHR) data. Deep learning models are leveraged to extract quantitative features from CT scans. These features combined with those directly read from the EHR database are fed into machine learning models to eventually output the probabilities of patient outcomes. This work demonstrates both the ability to apply a broad set of deep learning methods for general quantification of Chest CT scans and the ability to link these quantitative metrics to patient outcomes. The effectiveness of the proposed method is shown by testing it on an internally curated dataset, achieving a mean area under Receiver operating characteristic curve (AUC) of 0.77 on ICU admission prediction and a mean AUC of 0.73 on death prediction using the best performing classifiers. 
\end{abstract}
\begin{document}

\flushbottom
\maketitle
%
%
\thispagestyle{empty}

\section*{Introduction}
 Coronavirus Disease 2019 (COVID-19) was declared as a global pandemic by the World Health Organization in March 2020 and has resulted in millions of deaths globally \cite{world2021covid}. Of all the patients hospitalized due to COVID-19, 14-30\% require Intensive Care Unit (ICU) support and 20-33\% face death \cite{lassau2021integrating}. In order to treat a person severely infected by COVID-19, a lot of hospital resources are required, such as an ICU bed with a ventilator and hospital staffing \cite{centers2020strategies, kobokovich2020ventilator}. The scale of the pandemic has  led to severe bed shortages, and depleted hospital resources globally \cite{sen2021closer}.  Hence, it is of paramount importance to carefully manage ICU beds for effective management of the crisis. 

In order to tackle this, during the first few days of the onset of the disease, the ability to estimate the risk of fatality for each patient, would greatly assist hospitals in prioritising and treating the patients at high risk.
Leveraging machine learning, this work takes a step in this direction. Considering various factors related to COVID-19 as inputs to a machine learning model two predictions are made:  probability of a patient requiring an ICU bed and the probability of in-hospital mortality. In this work, we demonstrate the ability to quantitatively analyse images automatically to extract a subset of the inputs to machine learning models  and understand the influence of these features on patient outcomes.

While there are existing works on patient outcomes prediction, many of these works only rely on EHR data \cite{jimenez2021developing, estiri2021individualized}. A few other recently published papers dealing with image modality do not fully utilize the power of deep learning to extract various important metrics. For instance, Soda et al.\cite{soda2021aiforcovid} suggested using a convolutional neural network to extract features from a chest X-ray and directly uses it for prediction. Image features extracted in this fashion do not provide interpretability of the patient outcomes. Similarly, Zhang et al.\cite{zhang2020clinically} only used the information obtained from lesion segmentation. On the other hand, the proposed model uses multiple features extracted from the image modality besides including EHR metrics which act as inputs to the patient outcomes prediction model. These include a mixture of simple factual features like age and sex  extracted from Electronic Health Records (EHR) data; and features extracted using deep learning (DL) methods like normal lung percentage and muscle-and-fat ratio as a surrogate of frailty \cite{bunnell2021body}. 

All the DL methods that are used in this work are segmentation algorithms as we aim to quantify/delineate either anatomy or pathology. Hence, for extracting the various features using deep learning methods two architectures are mainly used: U-Net \cite{ronneberger2015u} and COPLE-Net \cite{wang2020noise}. The former architecture is based on `fully convolutional network' \cite{long2015fully} and builds upon it by making the upsampling network mostly symmetric to the downsampling network. On the other hand, COPLE-Net which was recently introduced for the segmentation of pneumonia lesions in COVID-19 patients improves upon U-Net and combines it with a noise-robust Dice loss function and a noise-robust learning framework in which a standard model (a.k.a. student) is guided by an Exponential Moving Average (EMA, a.k.a. teacher) of the model \cite{wang2020noise}.  Therefore, we leverage what is considered  the current standard in segmentation (UNet) along with recent methods when needed.

The contributions of this paper are two-fold: 1)  We illustrate how multiple deep learning architectures can be applied together to provide general quantification to extract quantitative metrics and provide an overall assessment of the patient state based on lung CT scans, and  2) We propose a novel multi-modal learning framework by collating the features extracted from CT scans and EHR data to predict the probability of two patient outcomes, namely admission to ICU and death. 

The rest of the paper is organised as follows: 
\begin{itemize}
    \item Firstly, an overview of the methods used in this work is provided. This is done by discussing the extraction of image-features followed by the classifiers used to predict the patient outcomes. Later, the implementation details are discussed which include the model training specifics such as hyper-parameters.
    \item Secondly, an overview of the datasets used in the framework is provided. This is done by first providing  a description of the datasets used for image-features extraction, and later by describing the dataset used for training the ML classifiers for patient outcomes modelling.
    \item Thirdly, quantitative and qualitative results are shown for all the DL models used for image-feature quantification followed by an evaluation of patient outcomes prediction.
    \item Finally, a summary of the framework is provided along with the contributions and future work.
\end{itemize}

\begin{figure}
    \centering
    \includegraphics{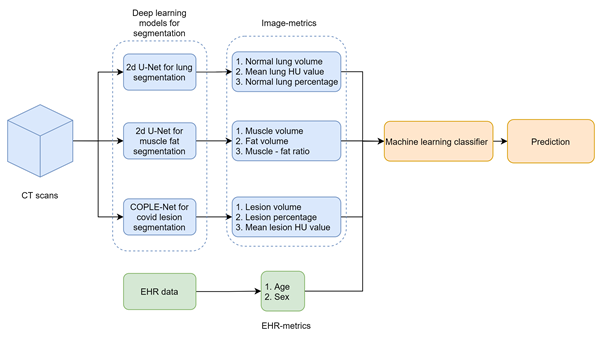}
    \caption{Overview of the proposed method. Various image-features are extracted using CT scans as inputs to multiple deep learning models. All the components related to image-features are shown in blue. These are coupled with features directly extracted from EHR. Components pertinent to EHR-features are shown in green.  All these features are then fed to a machine learning classifier to predict patient outcomes.}
    \label{fig:architecture}
\end{figure}

\section*{Methods}
In this section, the framework used to predict patient outcomes is described. Firstly, the inputs that are used to predict the outcomes is explained. Specifically, the ML classifier which predicts the patient outcomes, takes two types of inputs: EHR-features and image-features. The former is directly read from the EHR database, while the latter requires DL architectures for its computation and is discussed in detail. Secondly,  various machine learning (ML) classifiers that are used to model the outcomes are discussed in detail.

\subsection*{Extraction of image-based features}\label{ex_img_ftrs}
Image-related features are extracted using DL-based quantification of CT scans. The inputs for these DL models are the CT scans of the patients. Specifically, three DL networks are used to perform lung segmentation, ground glass and consolidation segmentation, and muscle-fat segmentation. Next, each of these networks is described along with the metrics extracted using the three architectures.

\subsubsection*{Lung segmentation metrics}
 For automatic lung segmentation, 2D U-Net \cite{ronneberger2015u} architecture is used, which has been shown to be effective for the task of lung segmentation \cite{skourt2018lung, shaziya2018automatic}. U-Net is based on the FCN architecture  \cite{long2015fully}. It primarily consists of a downsampling and an upsampling path. The former is a standard convolutional network made up by repeated application of two 3 $\times$ 3 convolutions and a 2 $\times$ 2 max pooling operation for downsampling. On the other hand, the upsampling path employs a transpose convolution and concatenates the resultant features at every step with the corresponding feature map from the downsampling path. At the final layer, a 1 $\times$ 1 convolution is used to map each feature vector to one of the two output classes (whether it is a part of the lung or not).

Using lung segmentation three metrics are extracted:
\begin{itemize}
    \item Normal lung volume (NL): The volume (in litres) of the lung in range -900HU to -700HU 
    \item Mean lung HU value (MCT): Mean Hounsfield unit (HU) within the lungs. 
    \item Normal lung percentage (NLperc): Fraction of normal lung to whole lung.
\end{itemize}

\subsubsection*{Muscle – Fat metrics}
Muscle-fat segmentation is 3-class segmentation problem and segments the image into one of the three categories: background, muscle and fat. Inspired by the success of U-Net in various medical segmentation tasks including muscle segmentation in Magnetic resonance imaging (MRI) scans \cite{ding2020deep, kemnitz2020clinical}, a 2D U-Net \cite{ronneberger2015u} is used for automatic muscle and fat segmentation. 

Eventually, three metrics related to muscle and fat segmentation are extracted:

\begin{itemize}
    \item Muscle volume (Muscle): The volume of muscle in litres
    \item Fat volume (Fat): The volume of fat in litres
    \item Muscle – fat ratio (MF ratio): The ratio of muscle volume to fat volume. 
\end{itemize}

The rationale behind using muscle and fat metrics is that the quantity of both muscle and fat have previously been reported as indicators of patient outcomes in a variety of studies. One such meta analysis of a variety studies looked at muscle and fat quantification in lung transplant patients \cite{rozenberg2020clinical}. They investigated studies that identified how differences in skeletal muscle and adiposity between patients can impact factors such as time on mechanical ventilation, time to hospital discharge, and overall survival. In most studies analysed (six of eleven), low muscle mass was an indicator for poorer patient outcomes. This could be linked to the effect that muscle mass has on metabolism and physiological reserve, with diminished muscle volume negatively affecting both of these aspects \cite{singer2015frailty, singer2016frailty}. Other studies have shown that increased adipose tissues (subcutaneous, visceral, and inter-muscular) are indicative of poorer patient outcomes in COVID-19 patients specifically \cite{bunnell2021body}. Higher than average abdominal subcutaneous adipose tissue, visceral adipose tissue and inter-muscular adipose tissue have been shown to be more likely to be admitted to ICU or die following the contraction of COVID-19. These measures have been shown to be more reliable than Body Mass Index (BMI) for predicting patient outcomes, given the inability of BMI to differentiate between muscle and fat \cite{muller2016beyond}. Further to this, it has previously been shown that variability in body composition for those with high BMIs can have a large effect on clinical outcomes for a variety of diseases \cite{muller2016beyond}. Because of these factors, the ratio between muscle and fat are more descriptive of the patient’s body composition, and thus can be used to better predict outcomes.
\subsubsection*{Ground glass and consolidation metrics}
For quantifying COVID-19 lesions in CT scans, COPLE-Net \cite{wang2020noise} is used, which is the current state-of-the-art for ground glass and consolidation segmentation. COPLE-Net is based on U-Net framework \cite{ronneberger2015u} and adapts it to the COVID-19 lesion segmentation by making a few changes. Firstly, it uses a combination of max and average pooling to reduce information loss during downsampling, compared to using only single max-pooling which has higher information loss. Secondly, it uses a 1 $\times$ 1 convolutional layer to reduce the dimension of the encoder output before concatenating it with the decoder features. Thirdly, it uses Atrous Spatial Pyramid Pooling \cite{chen2017deeplab} for better segmentation. 

To deal with noise in lesion annotations, the authors propose a new noise-robust loss function which is a generalization of Dice loss function \cite{milletari2016v} and Mean Absolute Error loss function. Further, a self-ensembling framework consisting of a student-teacher framework is deployed to train the network. In the student-teacher framework, both the teacher and the student are made up of the same network structure with the teacher model being an exponential moving average of the student model. The contribution of the student model to the teacher model is suppressed when the former shows a poor performance.  Moreover, the supervision of the teacher on the student is suppressed when it has a lower performance than the student. The reader is encouraged to read the original paper \cite{wang2020noise} for further details.

Using automatic covid lesion segmentation, three metrics are extracted:
\begin{itemize}
    \item Lesion volume (GG\_volume): The volume of the lung which contains lesions
    \item Lesion percentage (GG\_frac): The percentage of lung which contains lesions
    \item Mean lesion HU value (GG\_MHU): The Hounsfield unit (HU) within the lesion region. This is used to extract the severity of the lesion. 
\end{itemize}

To sum it up, using the three DL architectures described above, 9 different image-features are extracted. Next, the ML classifiers used to predict patient outcomes are expounded in detail.
\subsection*{Patient outcomes prediction}
Besides the 9 different image-features, 2 features (age and sex) are directly read from the EHR data. These are referred to as EHR-features. Combining these with the image-metrics, gives a total of 11 features that are used as inputs to the ML classifiers.

Moreover, the EHR data also contains the metrics ‘fraction of inspired oxygen’ (FIO2) and 'oxygen saturation' (SpO2) that can be used as features to the classifiers to predict patient outcomes. However, FIO2 metric is missing for approximately 61\% of the selected patients in the EHR data and hence was not used for analysis.  On the other hand, SpO2 did not improve the prediction of patient outcomes. As a result, there are two features that are directly extracted from EHR database and used for outcomes prediction.

It is also important to note that, for calculating image-features, the scan on the date closest to the date on which the patient returned a COVID-19 positive result is taken. This provides us with as much data as possible compared to selecting only those scans taken before/after COVID-19 diagnosis. Such a selection aims to capture how the underlying pathology or early onset pathology influences the patient pathway i.e. death or admission to ICU. Moreover, the reasoning behind this selection of a closest scan date is in line with the motivation of prediction of patient outcomes i.e. predicting the probability of an ICU admission or a death as soon as a subject tests positive so that appropriate further steps can be taken. However, in cases where the COVID-19 positive result date is not available in EHR data, we consider the latest available scan of the patient.

Once all the features i.e. both image-based and EHR-based are extracted,  several established classification models are utilised to predict two patient outcomes, namely, ICU admission and death. Five different ML classifiers are used in our experiments: logistic regression, support vector machines (SVM)\cite{cortes1995support}, random forest \cite{ho1995random}, ada-boost \cite{schapire2013explaining} and XGBoost \cite{Chen:2016:XST:2939672.2939785}. An overview of the proposed method is shown in Figure \ref{fig:architecture}.

Next, the implementation details of all the DL architectures and different ML classifiers used for our experiments are elucidated.

\subsection*{Implementation details}
\subsubsection*{Lung segmentation}
  
The lung segmentation model is trained using a weighted average of the binary cross entropy loss and the dice loss (BCE-Dice) with a ratio of 8:2. The model is optimized using Adam optimizer. The initial learning rate is set at 0.01 and is reduced by 90\% if the validation loss plateaus for 100 iterations. 15\% of the training set is used for validation. 

\subsubsection*{Muscle-Fat Segmentation}
For muscle-fat segmentation, each volume is divided into N slices since a 2D-UNET is used, where N is the number of slices per volume. The slices are further augmented by zooming the image using a random factor between 0.1 and 0.2. Adam optimizer was used for training the model with the default learning rate 0.001. The model was trained for 50 epochs and 1030 steps per epoch (the total number of slices) with a batch size equal to 1, using categorical cross entropy.

After the lung segmentations masks are inferenced, the connected regions in the 3D volume are extracted. Only the slices with the largest connected region are included to reject outliers (limiting the prediction to the muscle and fat around the lung region only). Noting that the same procedure is applied as a pre-processing step for inference.

\subsubsection*{Ground glass and consolidation segmentation}
For fine-tuning COPLE-Net model, the MONAI \cite{monai_consortium_2020_5525502} implementation of COPLE-Net is used. The initial learning rate is set to 1e-6 and is halved every 1000 iterations. All the other hyper parameters are same as the ones used by the authors of COPLE-Net.

\subsubsection*{Patient outcomes prediction}
For training and testing the machine learning classifiers to predict the patient outcomes, leave-one-patient-out (LOPO) scheme is used i.e. a model is tested on a single patient after training on all the remaining patients. Hence, if the dataset has $n$ different patients, there would be $n$ different LOPO runs. The outputs from each of these runs are collated to evaluate the performance of the classifiers. This is considered as one experiment. All the classifiers are implemented using scikit-learn \cite{scikit-learn}. In each of the LOPO runs, a grid search is performed  on the hyper-parameters mentioned below. The parameters that are not mentioned below are defaulted to the values provided by scikit-learn. 
\begin{itemize}
    \item \textit{Logistic regression}: L1 penalty is used and a bias is added to the decision function.
    \item \textit{Support vector machine}: Radial basis function (rbf) kernel with a kernel coefficient of 0.1 is used.
    \item \textit{Random forest}: The number of trees in the random forest is set to 100 and Gini impurity is used to measure the quality of the splits.
    \item \textit{AdaBoost classifier}: A grid search is performed over the number of maximum estimators (50, 100) and learning rate (1, 1.7, 1.33, 1.5),
    \item \textit{XG boost classifier}: A grid search is performed over max depth (3, 5, 7, 9).
\end{itemize}
 While training the classifiers, the minority class is oversampled using SMOTE \cite{chawla2002smote}.

\section*{Materials}
In this section, the datasets that have been used to train and test the DL models for extracting image-metrics are described followed by the dataset that has been used to train and test the ML classifiers for patient outcomes prediction.

\subsection*{Datasets used to train DL models for image-based feature extraction}
\label{im_ftrs_materials}
Beginning with lung segmentation, the model has been trained using volumetric CT images aggregated from two sources: a) publicly available data provided as a part of the Lung CT Segmentation Challenge (LCTSC) on Cancer Imaging Archive \cite{yang2017data}; b) National Health Service UK (NHS) regulated lung volumetric High Resolution Computed Tomography (HRCT) images with confirmed pathology other than COVID.
For the former dataset, the original split provided by the creators is used to segregate the train and test sets. The train set consists of 36 patient scans, while the test set set consists of 24 scans.
For the latter dataset, lung regions are manually annotated by in-house Image Analysis and divided into training and test set. Test set is independent of the training set. Training set consists of a total of 28 volumetric down-sampled images, and test set comprises of 5 volumetric HRCT images.
During the training process, training set is used for training and evaluation, while test set is reserved for the final test.
Secondly, for training the muscle-fat segmentation model, a newly-created dataset of 20 annotated patient scans is used. The model is tested on 25 slices from 25 different patients. Thirdly, COPLE-Net is fine-tuned using a mixture of publicly available datasets \cite{dataset1, dataset2} and a privately curated dataset. The two public datasets contain CT scans of 9 patients and 100 axial slices. The private dataset contributes an extra 253 lung slices annotated with ground glass and consolidation for training and  46 annotated slices for testing the performance of the model.

\subsection*{Dataset used for patient outcomes analysis}
To train a classifier for patient outcomes prediction, a patient cohort consisting of 2710 patients who had scans taken between the years 2007 and 2020 was collected from  ChelWest hospitals. A patient can undergo multiple CT scans at different time points and so there can be many more scans than the number of patients. For each patient, there exists a corresponding record in EHR containing information such as age, sex, if the patient has been admitted to ICU, the date on which the patient tested positive for COVID (if the patient was diagnosed with the disease) and if the patient eventually died. The record also contains the CT scan of the patient. Hence, a patient can be associated with multiple records of CT imaging taken at different time-points. 

For training the classifier, only the cases which have been diagonised with COVID are chosen. Further, all the patients who do not have a scan in 2020 are removed.  Each of the remaining scans is then manually checked to ensure that the scans are of good quality and the lungs are fully visible. Of the 2710 patients manually reviewed, 244 are used for patient outcomes analysis with the remaining being discarded. As there can be multiple scans corresponding to a single patient, the scan that is closest to the date on which the patient has been diagnosed with COVID is selected. This scan is used to extract the image-related features. Eventually, the image-metrics are combined with the EHR metrics for the remnant 244 patients and passed on to the classifiers. 

Among the 244 patients used for outcomes analysis, the distribution of one of the features ‘sex’ with respect to the patient outcomes is shown in Table \ref{tab:dataset}. These 244 patients are aged between 18 and 95 with an average age of 66.5.

\begin{table*}[ht!]
\caption{Distribution of patient outcomes with respect to sex in our newly curated dataset used for patient outcomes prediction.}
\vspace{2pt}
\centering
\begin{tabular}{lc cccc}
\toprule
& \multicolumn{2}{c}{Death} & \multicolumn{2}{c}{ICU}  \\
\cmidrule(r){2-3}\cmidrule(l){4-5}
Sex & Survived & Not survived  & Not admitted & Admitted     \\
\midrule
Male  & 88 & 39 & 113 & 14 \\
Female & 199 & 17 & 107 & 10\\
\bottomrule
\end{tabular}
\label{tab:dataset}
\end{table*}

\section*{Results}
In this section, the DL architectures deployed for image-based quantification are evaluated followed by a performance evaluation of different ML classifiers used for patient outcomes prediction.
\subsection*{Validation of DL models used for image-based quantification}
In order to establish the credibility of the various DL models that were used to extract image-features, quantitative and qualitative analyses of the performance of each model is provided. The latter has been performed by a Biomedical Image Analyst.

\subsubsection*{Lung segmentation}
Efficiency of the model prediction is evaluated using  Dice similarity as shown in Equation \ref{eq:first} by comparing the model prediction and the segmentation created by humans (otherwise called ground truth annotations) of the test set provided as a part of Task06\_Lung of Decathlon dataset.  Overall, high performance was achieved with an average dice score of 0.97 $\pm$ 0.019 were found across all the 24 patient scans in the test set. 

\begin{equation}
Dice(R_a, R_b) = \frac{2 |R_a \cap R_b|}{|R_a| + |R_b|},
    \label{eq:first}
\end{equation}
where $R_a$ is the predicted segmentation and $R_b$ is the ground truth segmentation. Dice score varies between 0 and 1 with 1 being the value for the prediction matching the ground truth.

\begin{figure}
\centering
\begin{subfigure}{.33\textwidth}
  \centering
  \includegraphics[width=.7\linewidth]{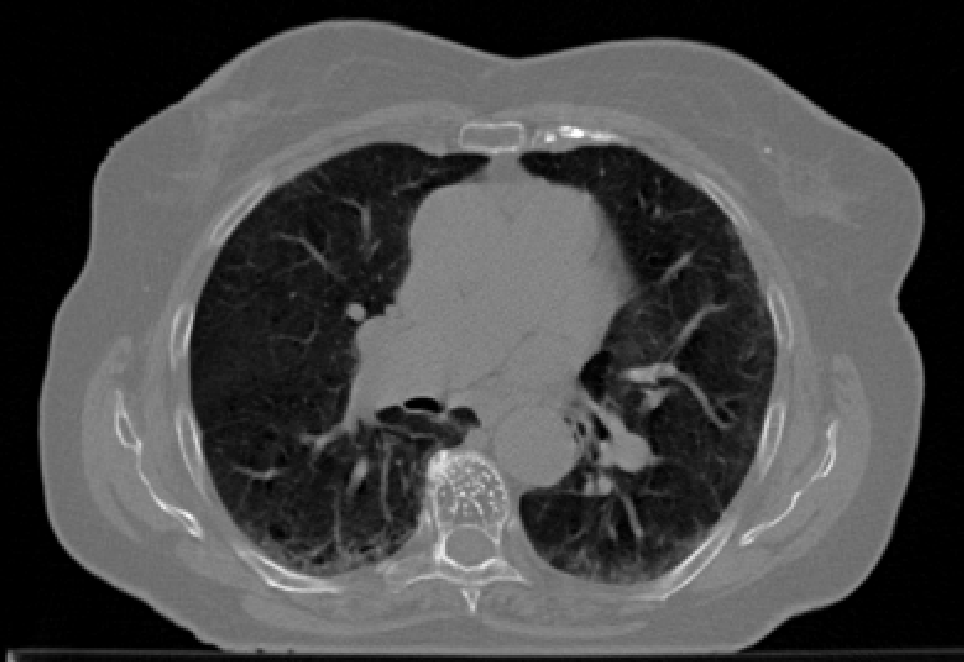}
  \caption{Raw scan taken from LCTSC \cite{yang2017data} }
  \label{fig:2}
\end{subfigure}%
\begin{subfigure}{.33\textwidth}
  \centering
  \includegraphics[width=.7\linewidth]{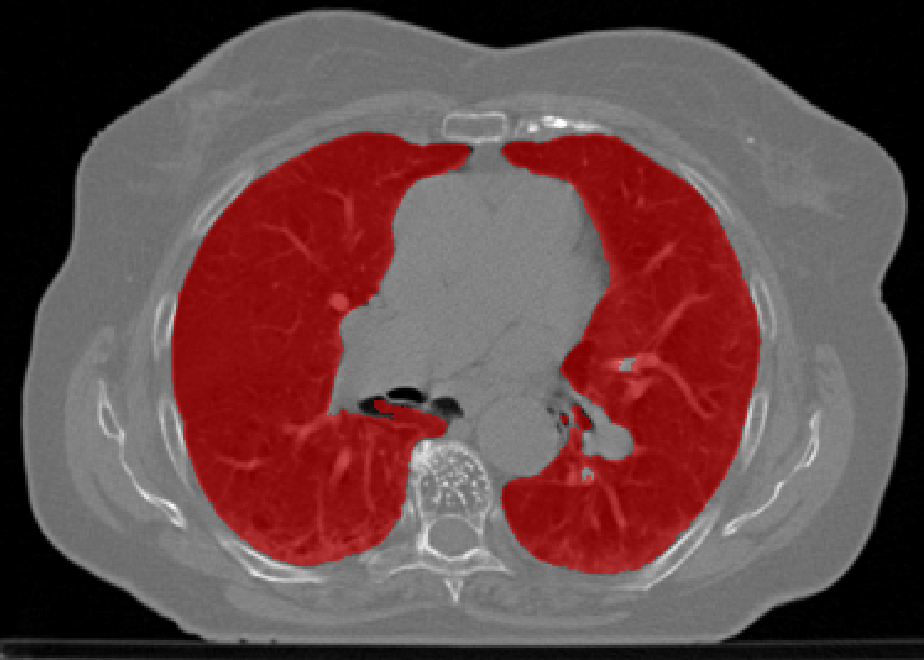}
  \caption{Predicted lung segmentation}
  \label{fig:lung2sub2}
\end{subfigure}
\begin{subfigure}{.33\textwidth}
  \centering
  \includegraphics[width=.7\linewidth]{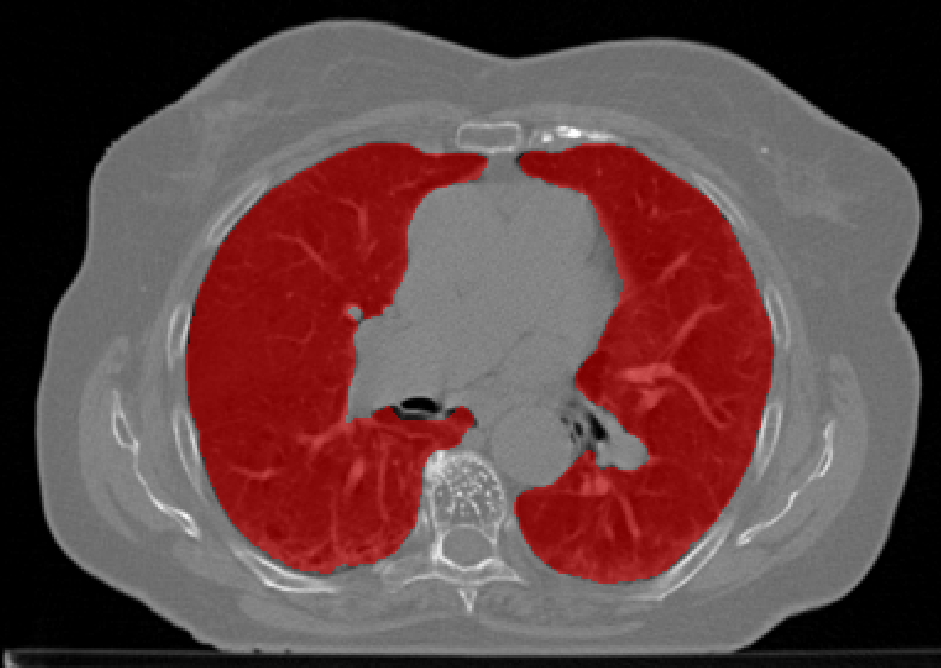}
  \caption{Ground truth lung segmentation}
  \label{fig:lungsub3}
\end{subfigure}
\caption{Raw scan along with predicted and ground truth (manually annotated) lung segmentations of CT scans taken from Lung CT Segmentation Challenge (LCTSC) \cite{yang2017data}}
\label{fig:lung_seg}
\end{figure}

Figure \ref{fig:lung_seg} shows two examples of the predicted lung segmentations taken from the test split of the Decathlon dataset. The lung segmentation model is able to delineate the boundaries of the thoracic cavity readily, while also including most pathological tissue in its segmentation. In cases with severe disease or large areas of pleural effusion, the accuracy of the model can drop slightly, although these are only in the most severe cases, presenting highly hyper/hypo intense areas of tissue. The model is also able to successfully remove the major hilar stuctures from the segmentation, resulting in a segmentation that focuses more specifically on the lung tissue itself, rather than supporting tissues such as major bronchi and blood vessels.

\begin{figure}
\centering
\begin{subfigure}{.33\textwidth}
  \centering
  \includegraphics[width=.7\linewidth]{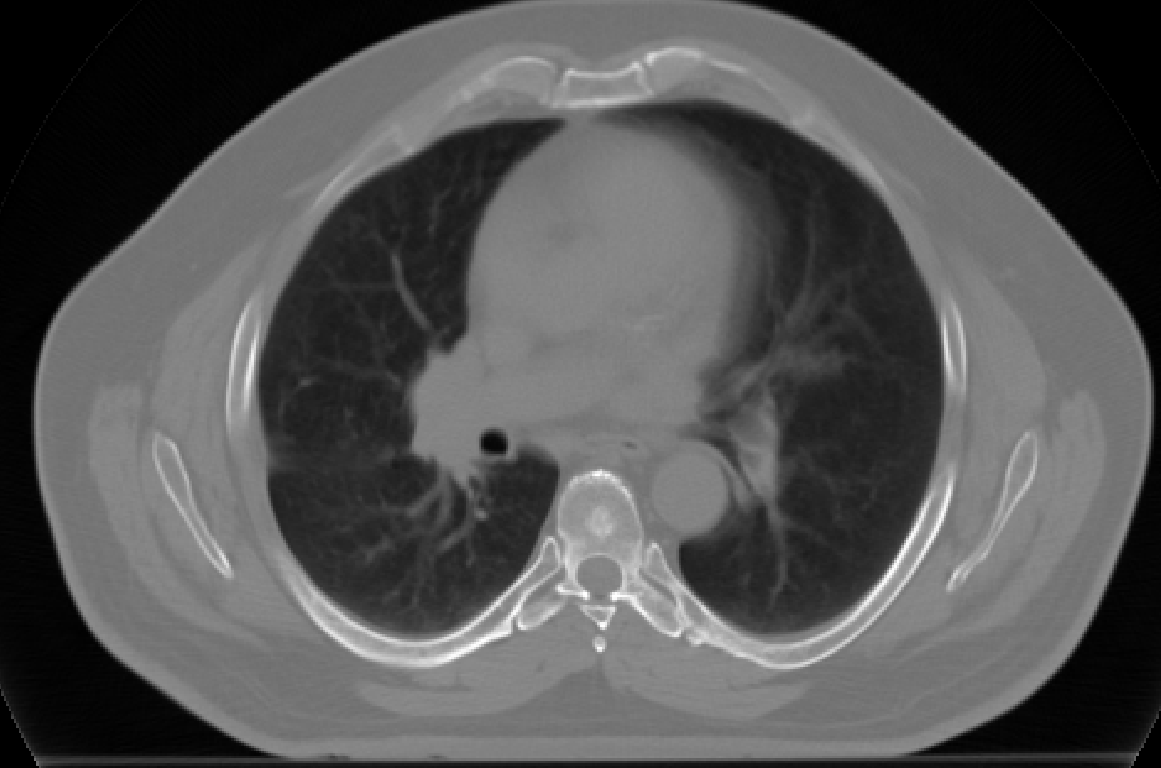}
  \caption{Raw scan taken from LCTSC \cite{yang2017data}}
  \label{fig:lung1sub1}
\end{subfigure}%
\begin{subfigure}{.33\textwidth}
  \centering
  \includegraphics[width=.7\linewidth]{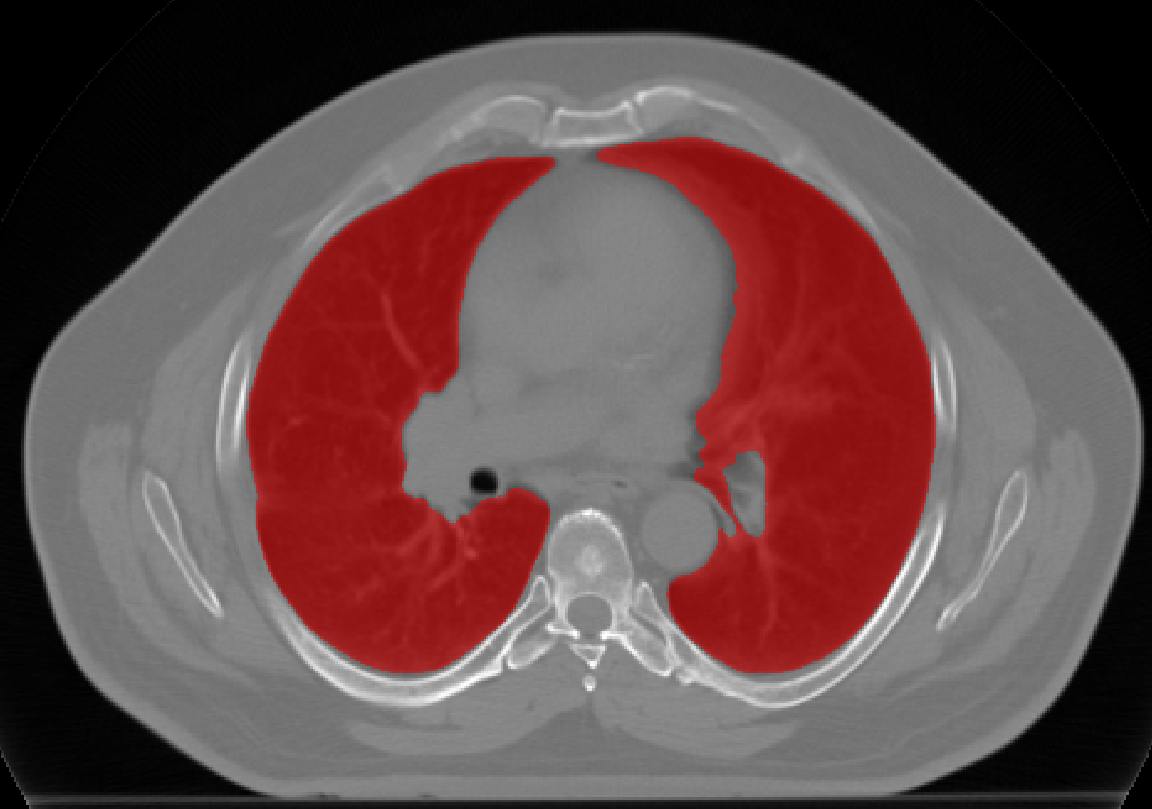}
  \caption{Predicted lung segmentation}
  \label{fig:lungseg}
\end{subfigure}
\begin{subfigure}{.33\textwidth}
  \centering
  \includegraphics[width=.7\linewidth]{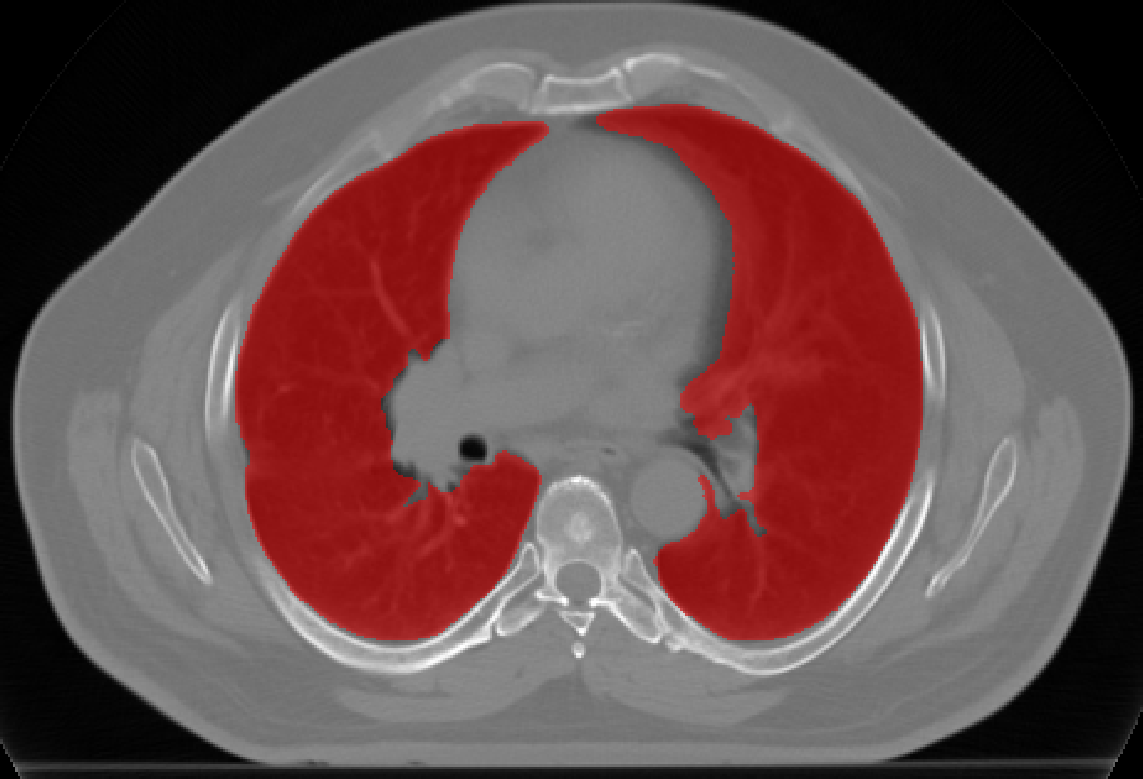}
  \caption{Ground truth lung segmentation}
  \label{fig:lungsub31}
\end{subfigure}
\caption{Raw scan along with predicted and ground truth (manually annotated) lung segmentations of CT scans taken from Lung CT Segmentation Challenge (LCTSC) \cite{yang2017data}}
\end{figure}

\subsubsection*{Muscle-Fat segmentation}
Similar to the lung segmentation, Dice score is used to evaluate the model. The predicted muscle-fat segmentations on the test set used for this purpose (containing the 25 slices as mentioned in the \textit{Materials} section) are compared to the ground truth annotations. As muscle-fat segmentation is 3-class segmentation problem, for each of the segmented slices, the dice score is averaged for each of the two foreground classes i.e. muscle and fat. Finally, an average of all the slices is evaluated which comes out to be 0.81. In a similar fashion, the model is also evaluated using Relative Volume Error (RVE) as shown in Equation \ref{eq:second}. 

\begin{equation}
    RVE(R_a, R_b) = \frac{abs(|R_a| - |R_b|)}{R_b},
    \label{eq:second}
\end{equation},
where $R_a$ is the predicted segmentation and $R_b$ is the ground truth segmentation. For RVE, a value of 0 means a perfect model with increasing values of RVE suggesting worse predictions. On the task of muscle-fat segmentation, the model achieves an RVE of 0.12.

\begin{figure}
\centering
\begin{subfigure}{.33\textwidth}
  \centering
  \includegraphics[width=.65\linewidth]{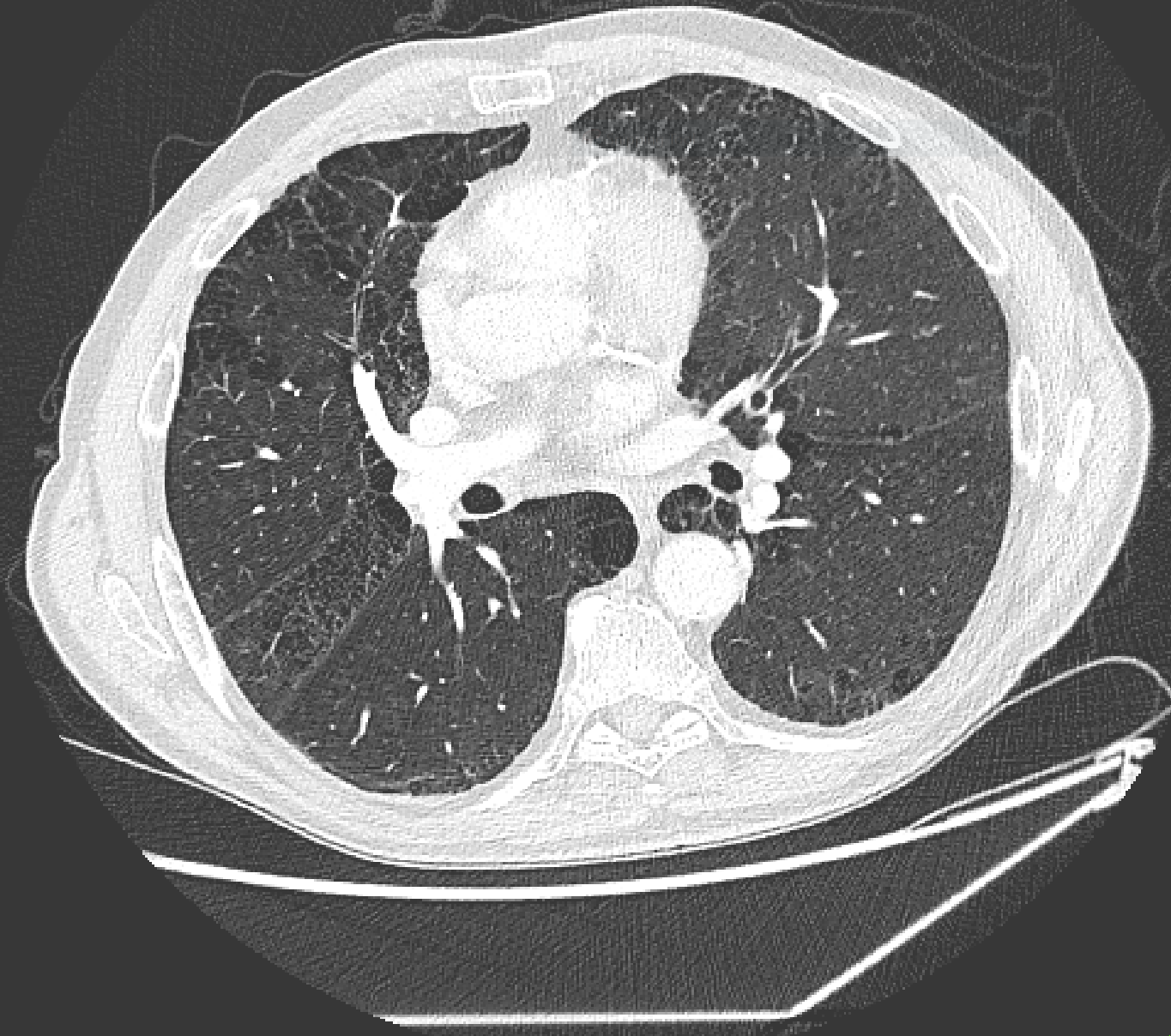}
  \caption{Raw scan taken from Decathlon \cite{antonelli2021medical} \\ dataset}
  \label{fig:mfat1sub3}
\end{subfigure}%
\begin{subfigure}{.33\textwidth}
  \centering
  \includegraphics[width=.65\linewidth]{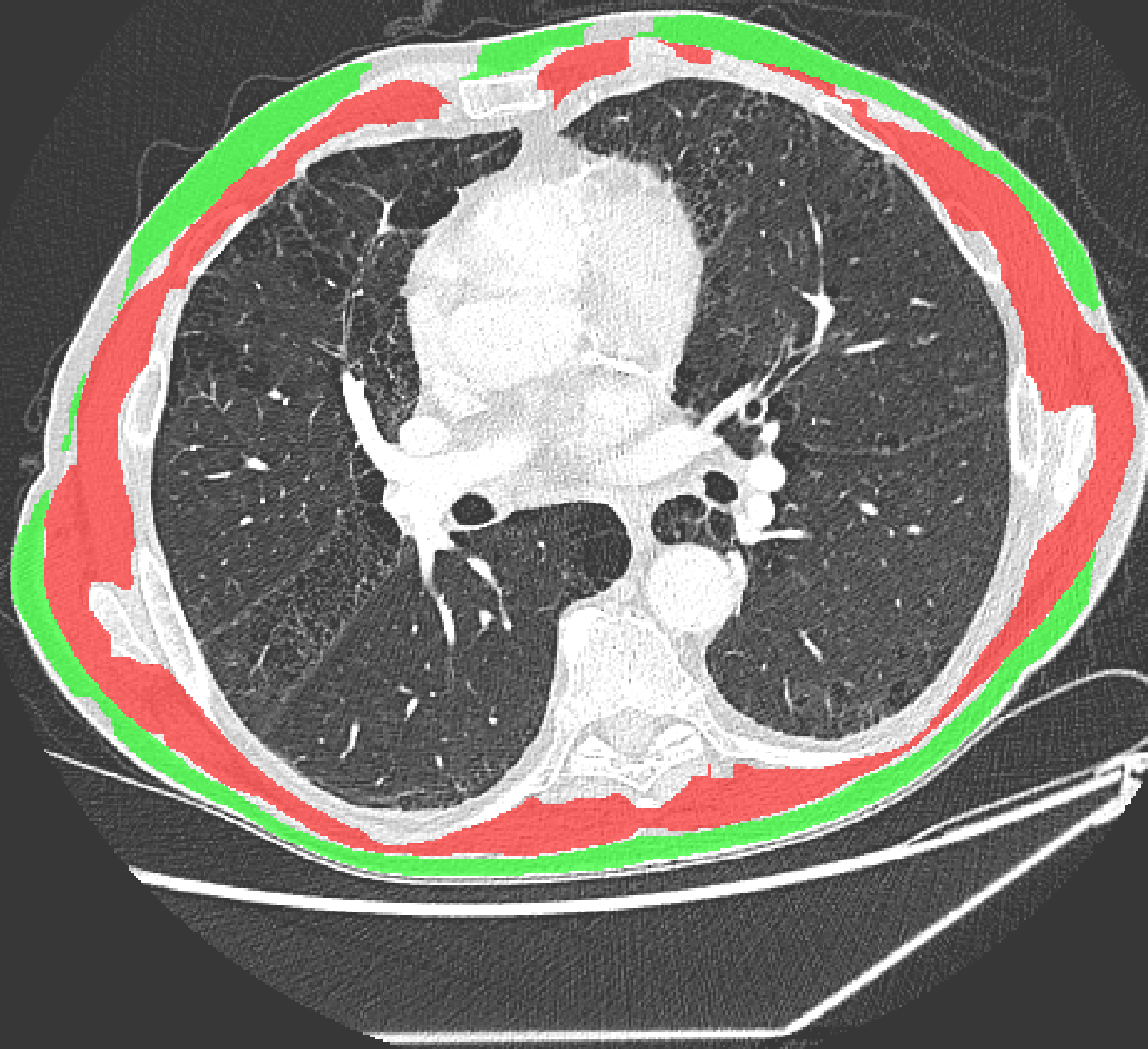}
  \caption{Predicted muscle-fat segmentation of an image taken from Decathlon \cite{antonelli2021medical}  dataset}
  \label{fig:mfat1sub4}
\end{subfigure}
\begin{subfigure}{.33\textwidth}
  \centering
  \includegraphics[width=.65\linewidth]{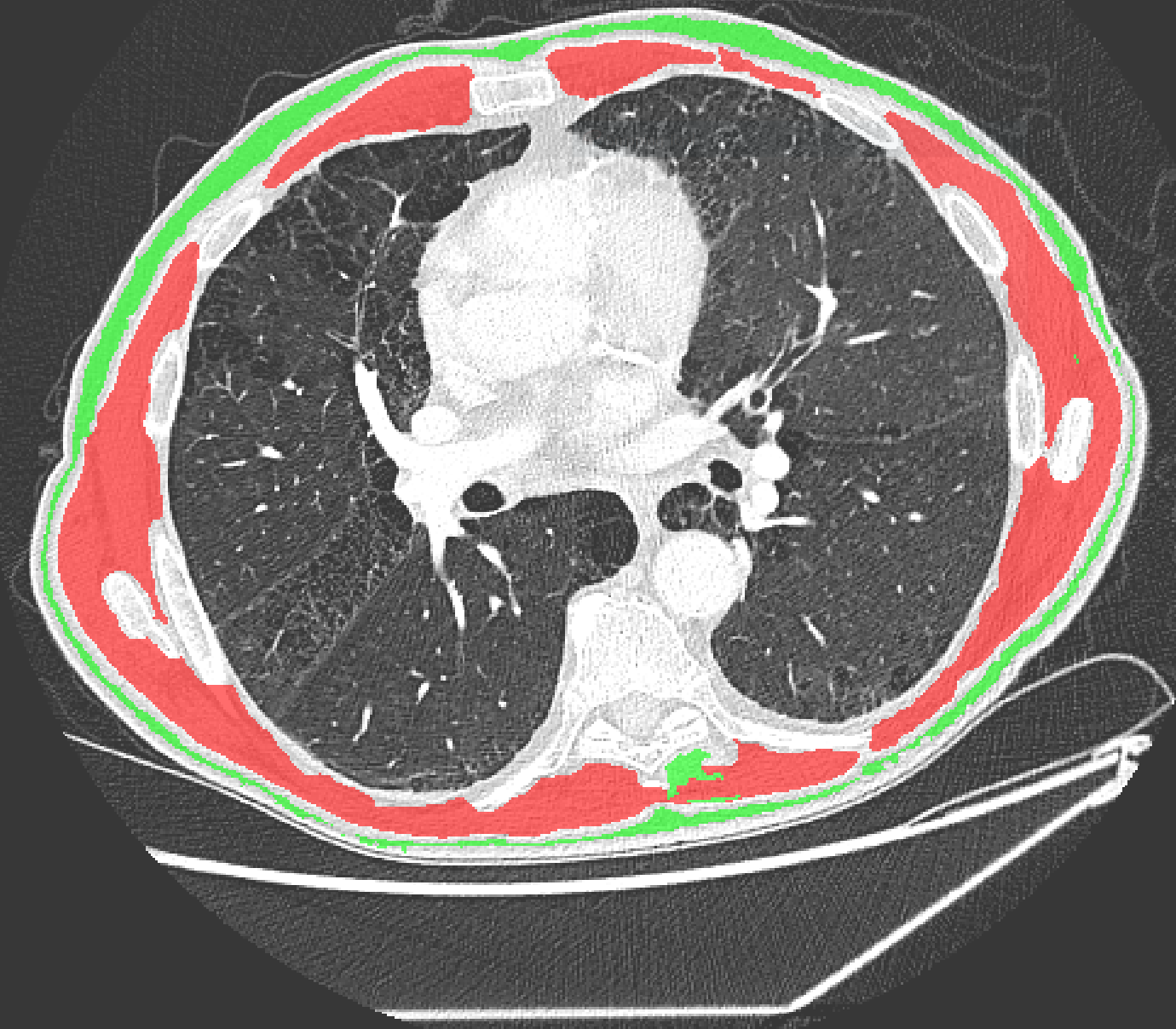}
  \caption{Ground truth muscle-fat segmentation of an image taken from Decathlon \cite{antonelli2021medical} dataset}
  \label{fig:mfat2sub4}
\end{subfigure}
\caption{Raw scan along with predicted and ground truth (manually annotated) muscle-fat segmentations of a lung slice taken from Decathlon dataset}
\label{fig:mfat1}
\end{figure}

\begin{figure}
\centering
\begin{subfigure}{.33\textwidth}
  \centering
  \includegraphics[width=.65\linewidth]{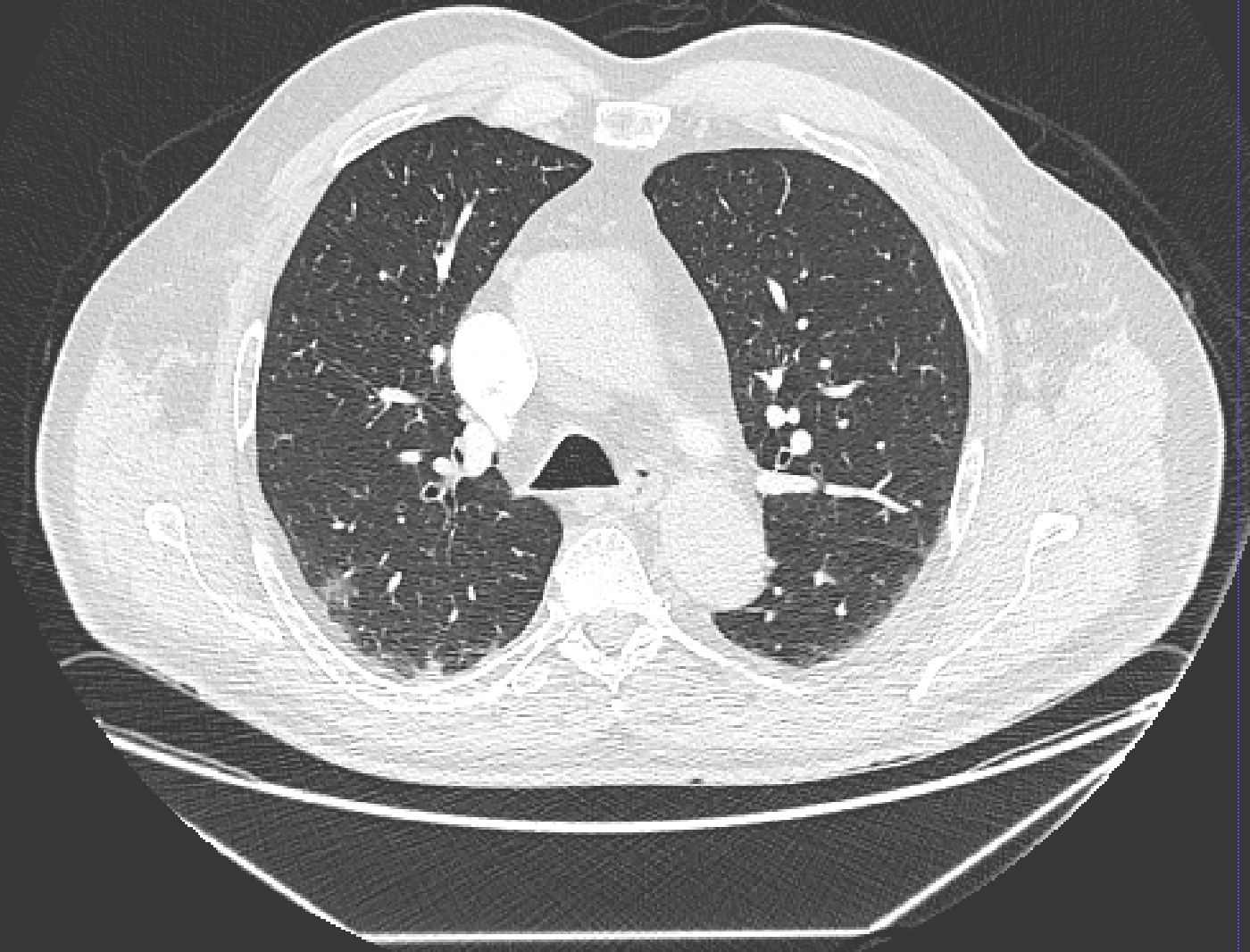}
  \caption{Raw scan taken from Decathlon \cite{antonelli2021medical} \\ dataset}
\end{subfigure}%
\begin{subfigure}{.33\textwidth}
  \centering
  \includegraphics[width=.65\linewidth]{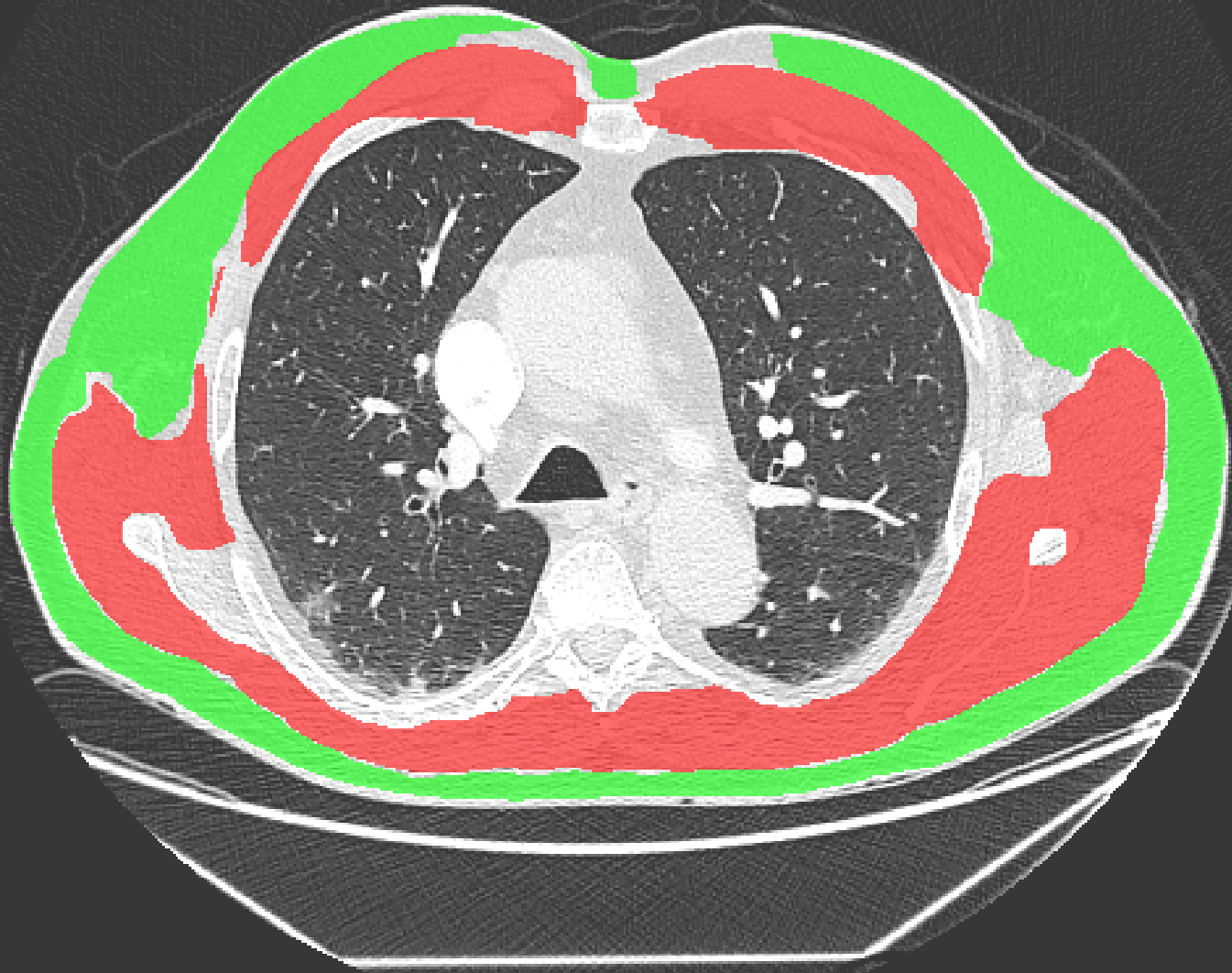}
  \caption{Predicted muscle-fat segmentation of an image taken from Decathlon \cite{antonelli2021medical} datase4}
\end{subfigure}
\begin{subfigure}{.33\textwidth}
  \centering
  \includegraphics[width=.65\linewidth]{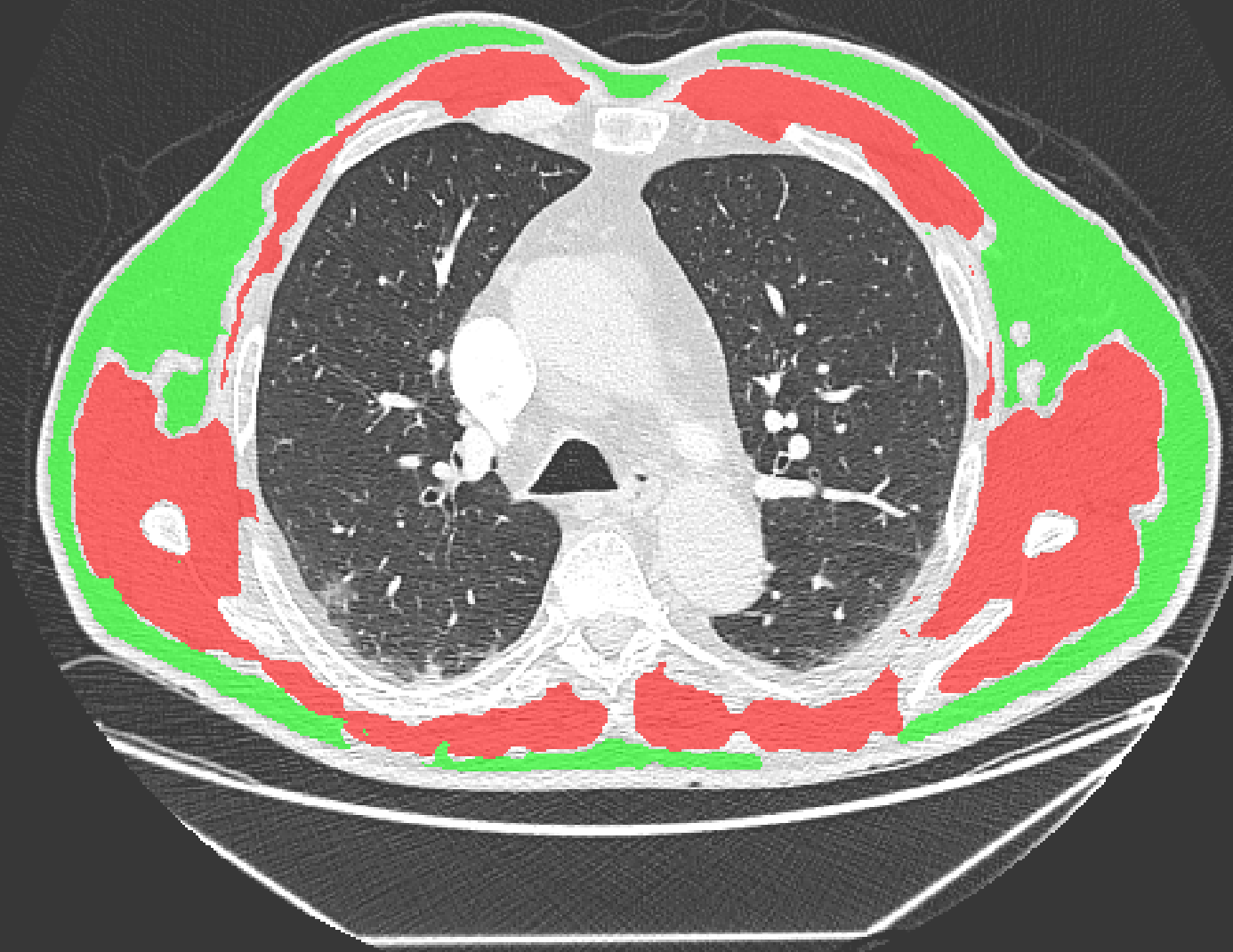}
  \caption{Ground truth muscle-fat segmentation of an image taken from Decathlon \cite{antonelli2021medical} dataset}
  
\end{subfigure}
\caption{Raw, predicted and ground truth (manually annotated) muscle-fat segmentations of a lung slice taken from Decathlon  dataset}
\label{fig:mfat2}
\end{figure}

Two images (taken from Decathlon dataset for displaying in the paper) of the predicted muscle-fat segmentations are shown in Figures \ref{fig:mfat1} and \ref{fig:mfat2}. The model is able to identify and segment the skeletal muscle and peripheral fat of the thoracic and upper abdominal regions, while managing to ignore most bone and tissues within the body cavities themselves. There are some examples of decreased model performance, or cases where over/under-segmentation is present, however these cases are restricted to scans with low signal to noise ratios caused by artefacts that obscure the surrounding tissues.

\begin{figure}
\centering
\begin{subfigure}{.33\textwidth}
  \centering
  \includegraphics[width=.65\linewidth]{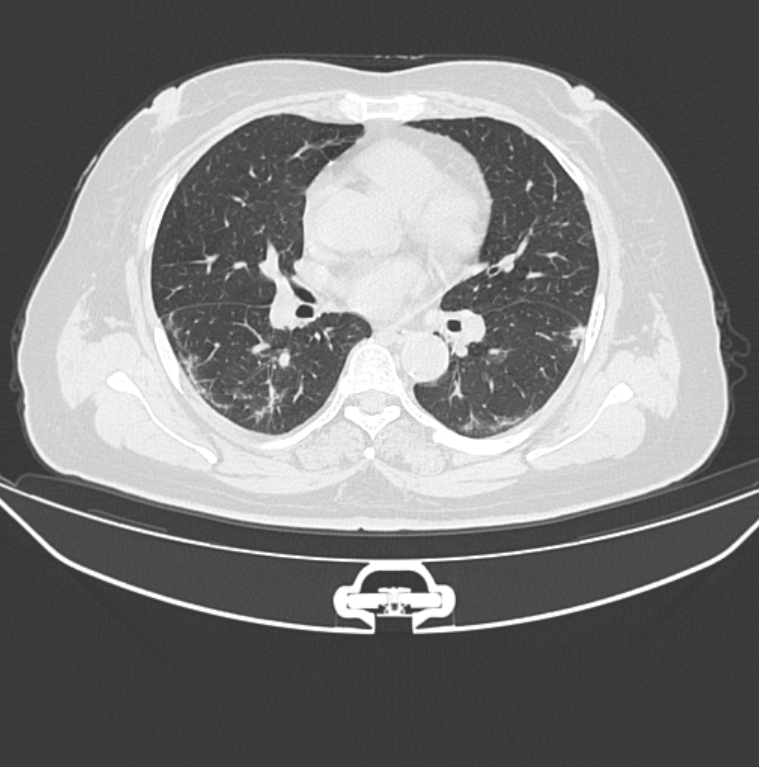}
  \caption{Raw scan taken from COVID-19 Lung \\ and Infection dataset \cite{ma_jun_2020_3757476}}
\end{subfigure}
\begin{subfigure}{.33\textwidth}
  \centering
  \includegraphics[width=.65\linewidth]{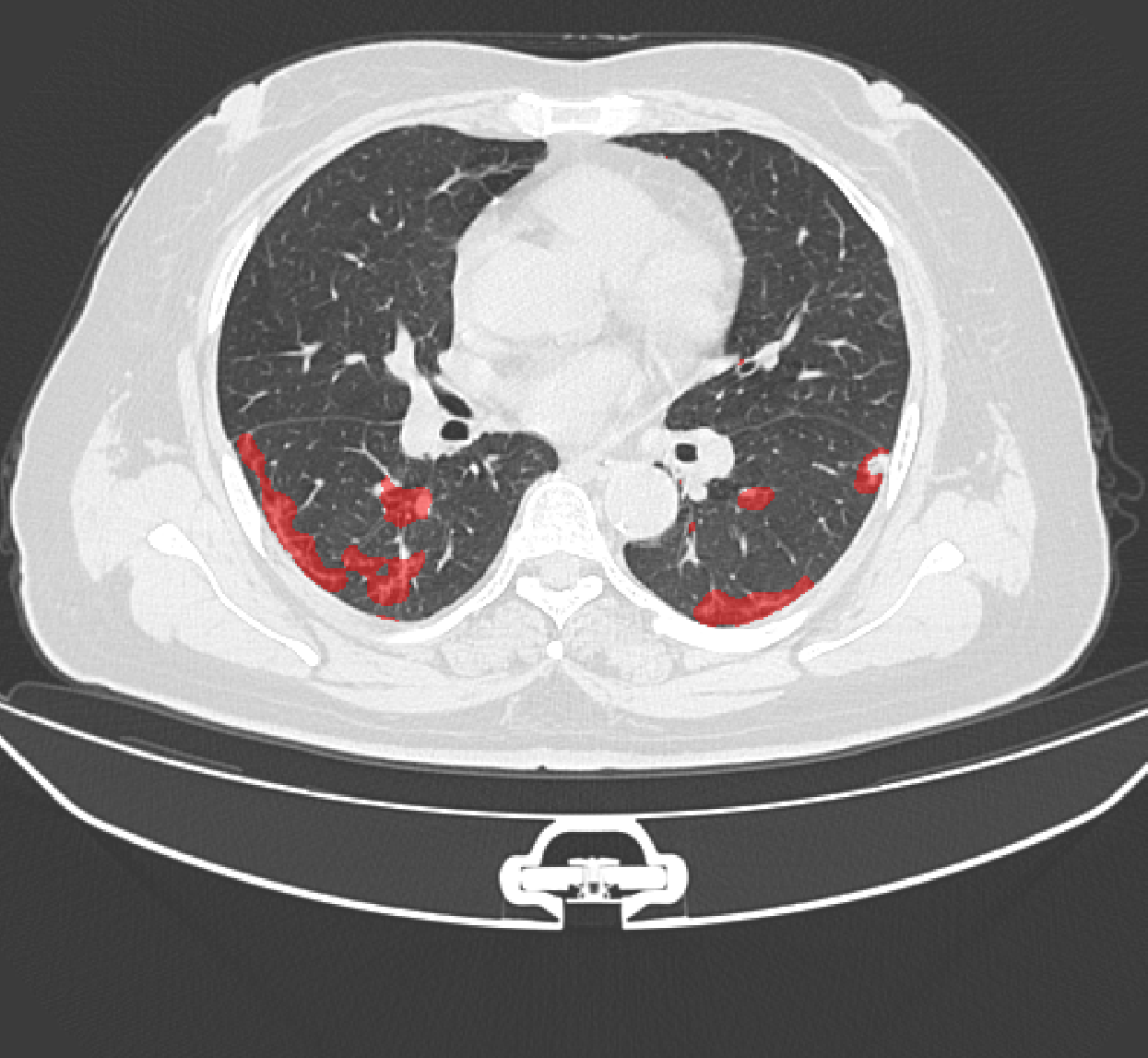}
  \caption{Predicted ground glass and consolidation segmentation of the scan}
  
\end{subfigure}
\begin{subfigure}{.33\textwidth}
  \centering
  \includegraphics[width=.65\linewidth]{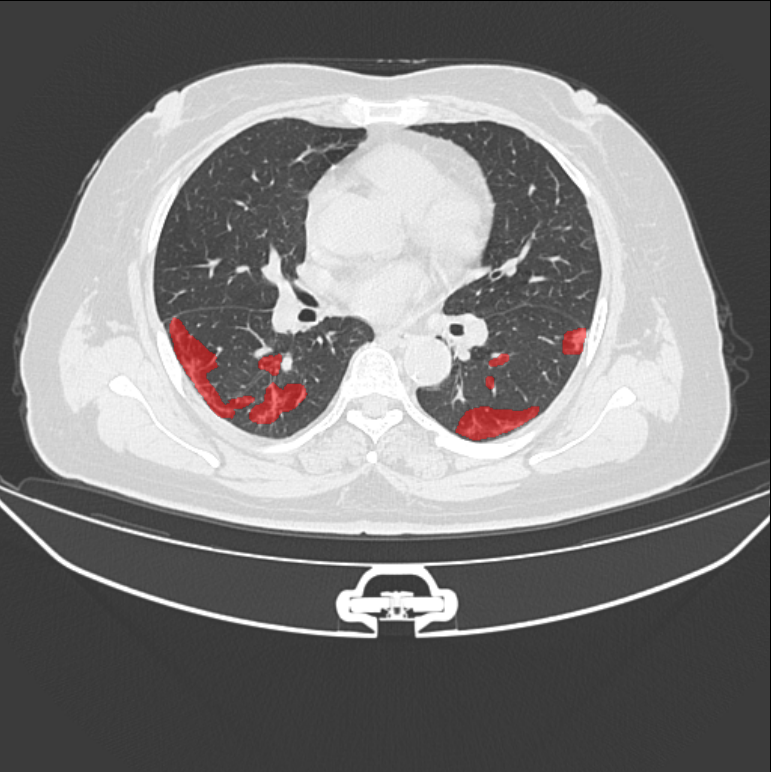}
  \caption{Ground truth segmentation of ground glass and consolidation of the scan}
  
\end{subfigure}
\caption{Raw scan along with predicted and ground truth (manually annotated) ground glass and consolidation segmentations of a lung slice from COVID-19 Lung and Infection dataset \cite{ma_jun_2020_3757476}}
\label{fig:gg1}
\end{figure}

\begin{figure}
\centering
\begin{subfigure}{.33\textwidth}
  \centering
  \includegraphics[width=.65\linewidth]{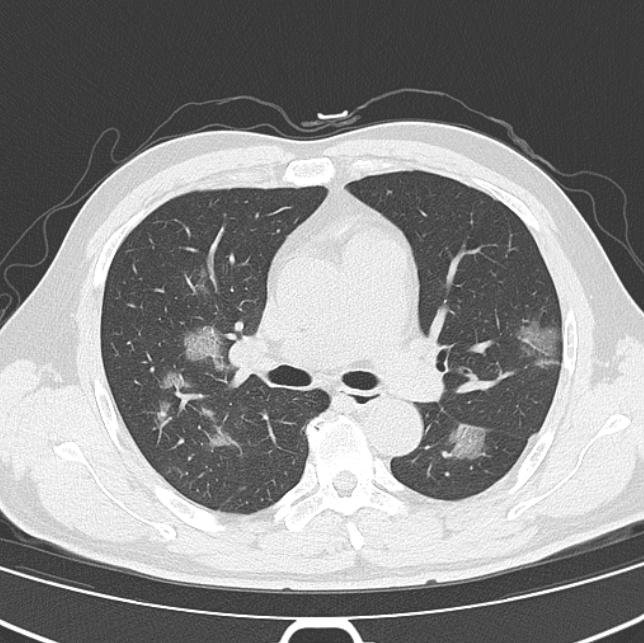}
  \caption{Raw scan taken from COVID-19 Lung \\ and Infection dataset \cite{ma_jun_2020_3757476} }
  \label{fig:sub3}
\end{subfigure}%
\begin{subfigure}{.33\textwidth}
  \centering
  \includegraphics[width=.65\linewidth]{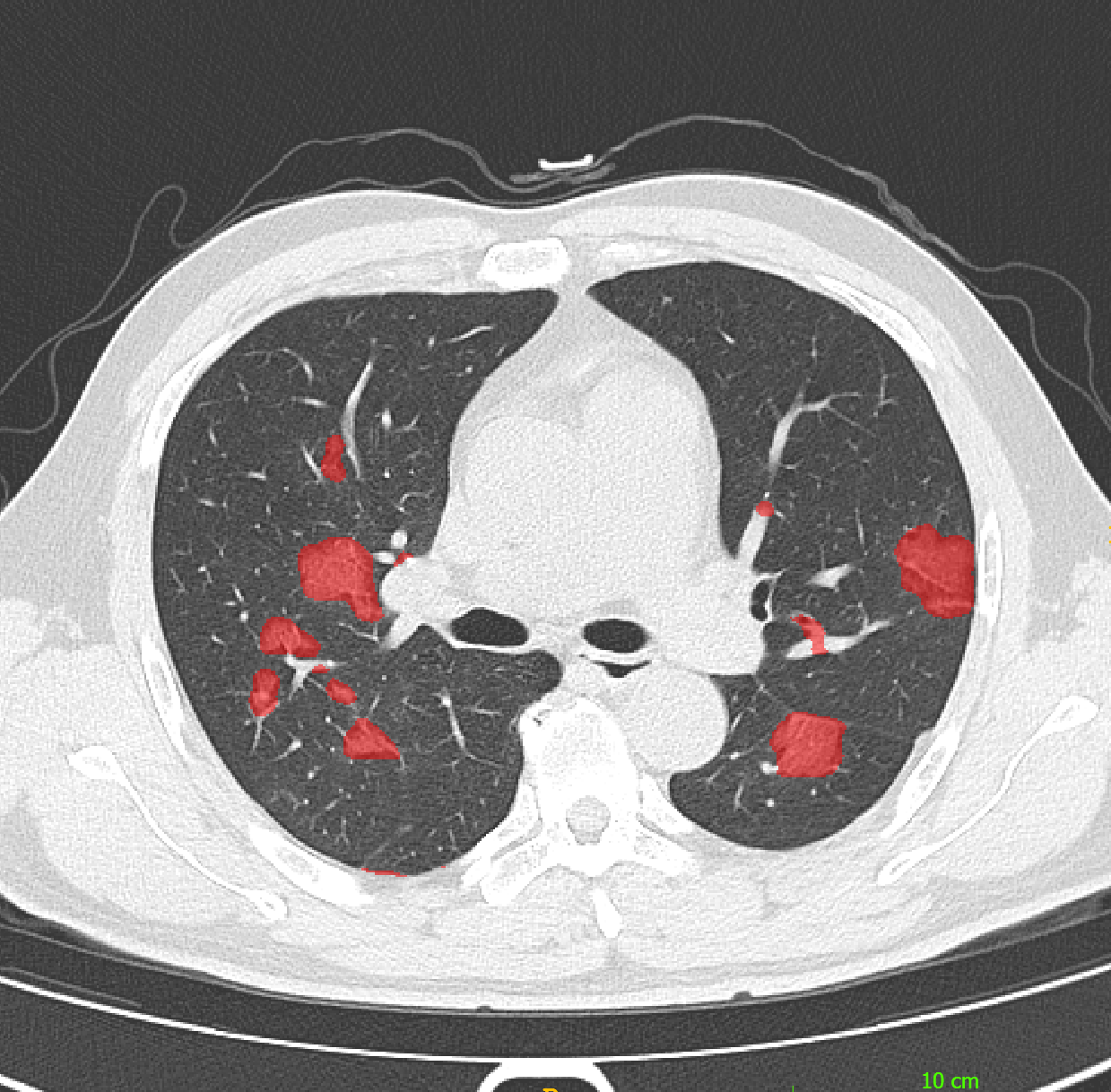}
  \caption{Predicted ground glass and consolidation segmentation of the scan}
  
\end{subfigure}
\begin{subfigure}{.33\textwidth}
  \centering
  \includegraphics[width=.65\linewidth]{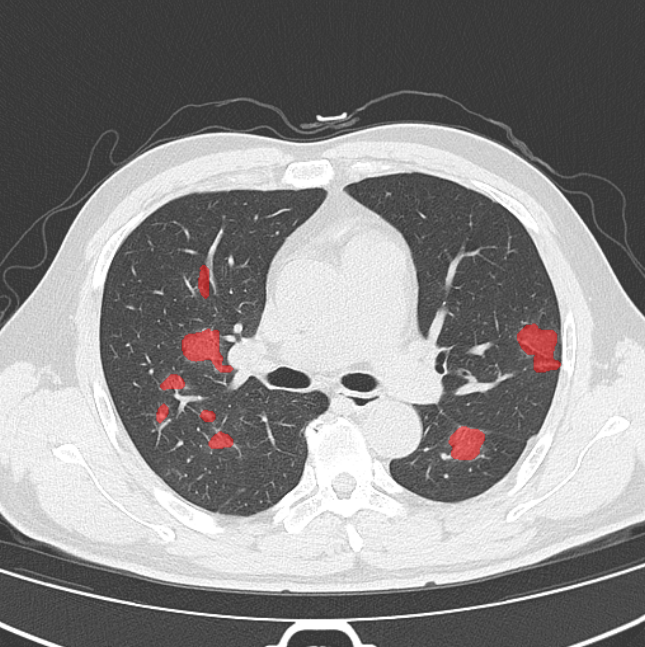}
  \caption{Ground truth segmentation of ground glass and consolidation of the scan}
  
\end{subfigure}
\caption{Raw scan along with predicted and ground truth (manually annotated) ground glass and consolidation segmentations of a lung slice from COVID-19 Lung and Infection dataset \cite{ma_jun_2020_3757476}}
\label{fig:gg2}
\end{figure}

\subsubsection*{Ground glass and consolidation segmentation}
For a quantitative evaluation of the fine-tuned COPLE-Net model, Dice similarity and RVE are used.

The predicted annotations on the test set (containing 46 slices as mentioned in the \textit{Materials} section) are compared to the ground truth data by using the Equations \ref{eq:first} and \ref{eq:second}. The model gives an average Dice score of 0.55 and an average relative volume error of 1.06

Figures \ref{fig:gg1} and \ref{fig:gg2} show images of predicted and ground truth segmentations of COVID-19 lesions (ground glass and consolidation). The model was seen to perform well, detecting areas of ground glass opacity (GGO) efficiently. The model is able to handle widespread areas of GGO as well as smaller foci showing it's flexibility. The accuracy of the model can be seen to dip when GGO is detected around blood vessels and other hyperintense structures, where the increased attenuation causes a slight local increase in the immediate surroundings. Despite the success of the model in handling GGO, in cases with severe consolidation the proficiency of the model dips. This occurs especially when the areas of consolidation are in close proximity to the thoracic wall. This issue is linked to the lung segmentation model however rather than a standalone issue with this model. This decrease in performance is not seen when the extent of consolidation is limited to smaller areas, which are picked up by the model as efficiently as GGO is with this model.

\begin{figure}
\centering
\begin{subfigure}{.5\textwidth}
  \centering
  \includegraphics[width=.7\linewidth]{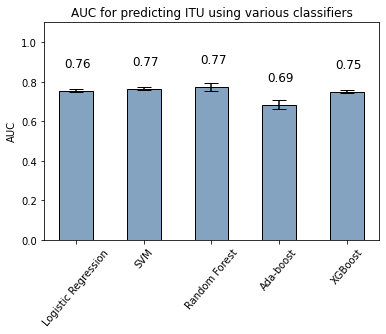}
  \caption{ROC curves of various classifiers for predicting ICU admission}
  \label{fig:sub1}
\end{subfigure}%
\begin{subfigure}{.5\textwidth}
  \centering
  \includegraphics[width=.7\linewidth]{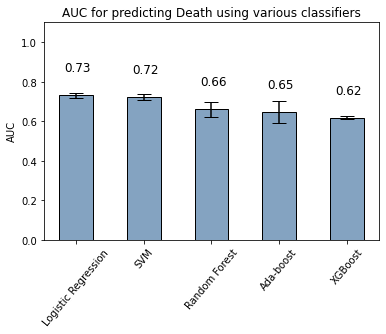}
  \caption{ROC curves of various classifiers for predicting death}
  \label{fig:rocsub2}
\end{subfigure}
\caption{ROC curves for various patient outcomes}
\label{fig:roc_curves}
\end{figure}

\subsection*{Patient outcomes prediction}
For assessing the performance of various ML classifiers used in modelling patient outcomes, the predictions from 244 different LOPO runs (as there are 244 patients in the dataset used for outcomes analysis) are collated and two different graphs, namely, the Receiver Operating Characteristic (ROC) curve and feature importance graph are plotted for each of the classifers. This is considered as one experiment. 

An ROC curve shows performance of a classification model at all classification thresholds and contains two parameters - True Positive Rate ($TPR$) and False Positive Rate ($FPR$), calculated as shown in Equations \ref{eq:tpr} and \ref{eq:fpr}.
\begin{equation}
TPR = \frac{TP}{TP + FN},
    \label{eq:tpr}
\end{equation}
where $TP$ and $FN$ represnt true positives and false negatives respectively.
\begin{equation}
    FPR = \frac{FP}{FP + TN},
    \label{eq:fpr}
\end{equation}
where $FP$ and $TN$ represnt false positives and true negatives respectively.

In order to account for the randomness in the classifiers, each experiment is run 10 different times and the mean (along with a 95\% confidence interval) area under the ROC curve (AUC) is calculated and shown for all the methods in Figure \ref{fig:roc_curves}. While Random Forest classifier with an AUC of 0.77, 95\% CI (0.75–0.79) ) and SVM with an AUC of 0.77, 95\% CI (0.76-0.77) perform the best in ICU admission prediction, Logistic Regression with a mean AUC of 0.73, 95\% CI(0.72 - 0.74) performs the best for death prediction.

\begin{figure}
\centering
\begin{subfigure}{.4\textwidth}
  \centering
  \includegraphics[width=.7\linewidth]{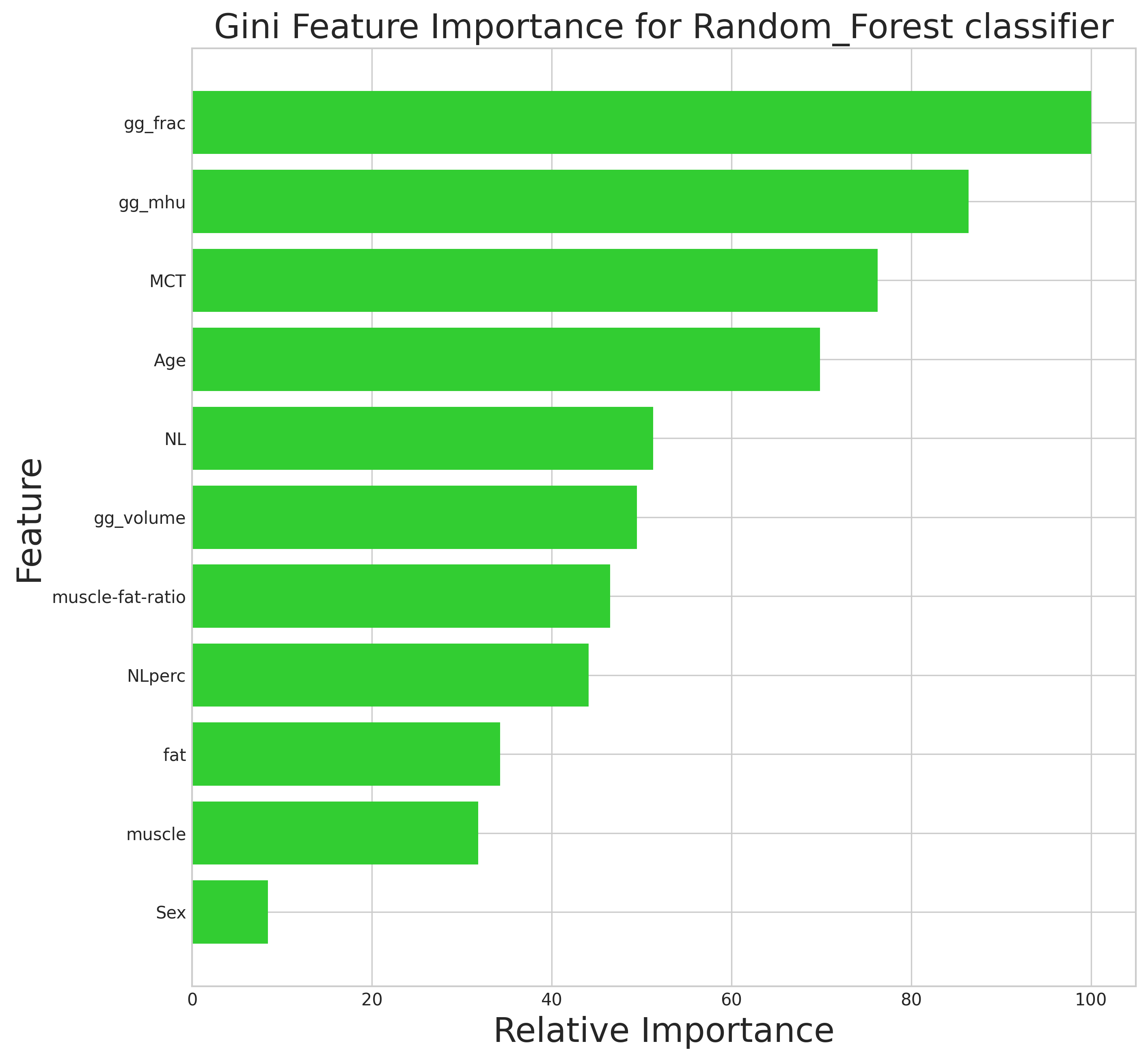}
  \caption{Importance of various features for predicting ICU admission}
  \label{fig:s}
\end{subfigure}
\begin{subfigure}{.4\textwidth}
  \centering
  \includegraphics[width=.7\linewidth]{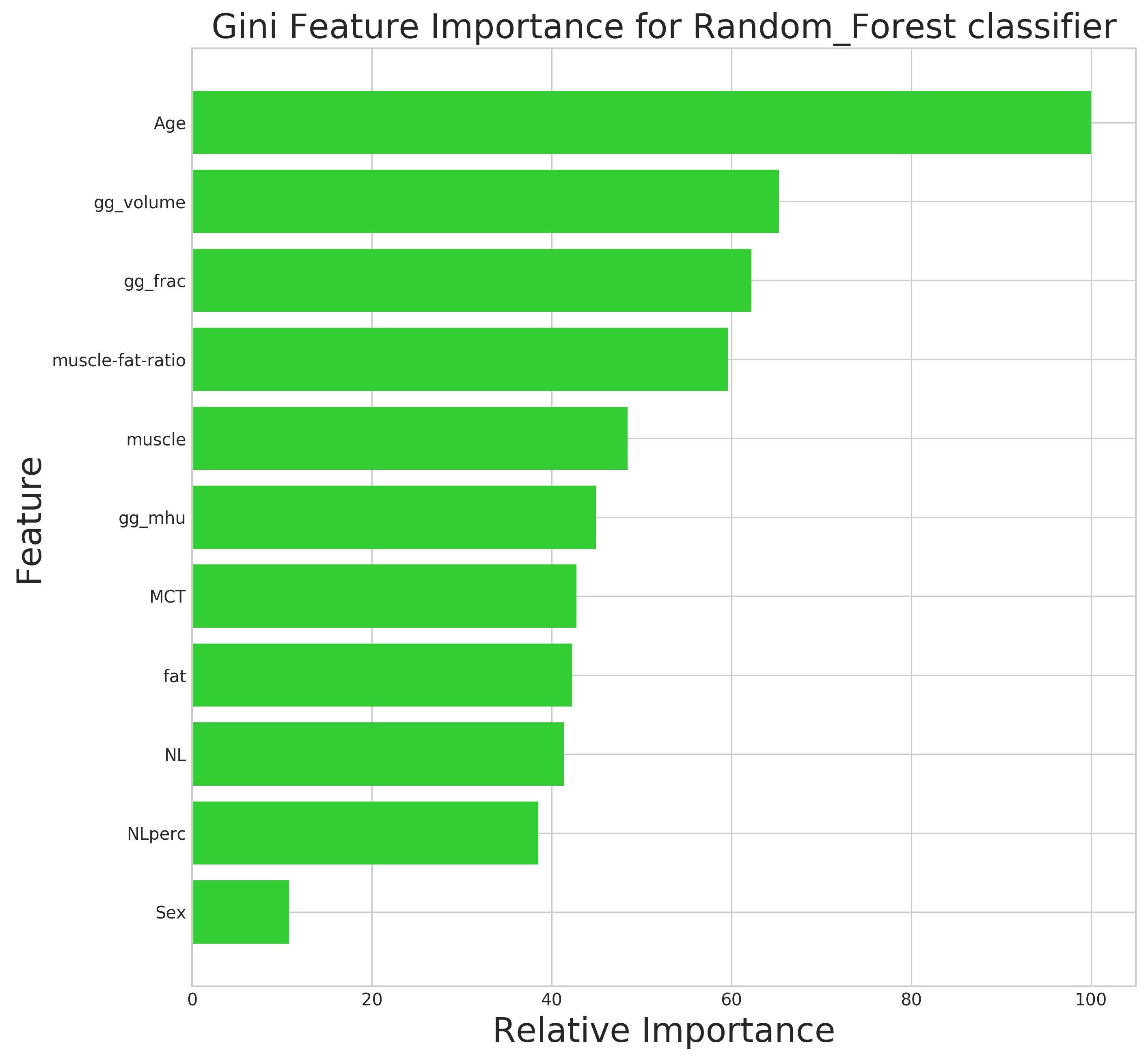}
  \caption{Importance of various features for predicting death}
  \label{fig:impsub2}
\end{subfigure}
\caption{Importance of various features for predicting patient outcomes using Random Forest classifier. Refer to  Methods section for the explanation of the features.}
\label{fig:imp_curves}
\end{figure}

Secondly, the Gini feature importance \cite{loh2011classification} is calculated for all the features when using the random forest classifier. Gini importance \cite{loh2011classification} for a given feature shows the influence of a given feature on the prediction and is calculated using Gini impurity \cite{loh2011classification}. For a given node, Gini impurity is calculated as shown in Equation \ref{eq:gini}

\begin{equation}
    \text{Gini impurity}  = \sum_{i=1}^{n}(p_i)(1 - p_i),
    \label{eq:gini}
\end{equation}
where $p_i$ is the frequency of the class $i$ at a given node and $n = 2$ for a binary classification. For calculating the importance of a given feature, the normalized sum of all impurity decrease values for nodes split on the given feature is considered. The weightage of each node is equivalent to the probability of reaching that node, which is approximated by the proportion of samples reaching that node. The feature importance  graph for the various features that were extracted is shown in Figure \ref{fig:imp_curves}. Feature importance graphs shows insights into the features which predominantly determine the patient outcomes. For instance, prediction of death is majorly determined by factors such as age, ground glass and consolidation volume, and fraction of lungs affected by ground glass and consolidation. Similarly, the major factors affecting the ICU admissions include fraction of lungs affected by ground glass and consolidation, mean HU within the part of the lung affectedby ground glass and consolidation. 

Finally, in order to show the relationship between the outcomes and the input features, correlation is  calculated and is shown in Table \ref{tab:corr}. A positive correlation implies that an increase in the corresponding input value is likely to result in an increased probability of observing the outcome under consideration and vice-versa. Moreover, the magnitude of the correlation coefficient signifies the strength with which the input and the outcome are correlated. We observe that NL, NLperc, muscle and MF ratio are inversely related to patient outcomes. For instance, higher NL correlates to lesser probability of death and ICU admission. Similarly, features reflecting abnormalities like GG\_frac and GG\_volume are directly proportional to outcomes. Moreover, higher MCT and GG\_MHU values reflect increased probability of death and ICU. This is because, generally, higher MCT values correspond to abnormalities in lungs. 

Two correlation co-effecients demand further attention. Firstly, the correlation between fat volume and outcomes is not what was expected from this study, given obesity is known to be a risk factor for worse patient outcomes in COVID-19 patients\cite{gao2021associations}.  Given this however, the data provided did not include BMI measurements or information on the patients weight. It is conceivable that the patients included in this study had a closer to average body composition, with little to no obese or morbidly obese patients included that may have swayed the body fat measurements towards more negative outcomes. Secondly, age was found to be negatively correlated to ICU admission. On further examination, it was found that out of the 92 patients aged above 76, 34 patients died while only 3 were admitted to ICU. This may have been due to the severity of disease upon admission.

\begin{table*}[ht!]
\caption{Correlation coefficient between various inputs and the outcomes}
\vspace{2pt}
\centering
\begin{tabular}{lc cc}
\toprule
Input & Death & ICU    \\
\midrule
Normal Lung volume (NL)  & -0.1 & -0.23\\
Mean lung HU value (MCT) &  0.14 & 0.35\\
Normal lung percentage (NLperc) & -0.13 & -0.28 \\
Muscle volume (Muscle) & -0.029 & -0.018 \\
Fat volume (Fat) & -0.11 & -0.043 \\
 Muscle – fat ratio (MF ratio)  & 0.053 & -0.021 \\
 Age & 0.25 & -0.15 \\
 Sex & -0.19 & -0.042 \\
 Lesion percentage (GG\_frac) & 0.19 & 0.28 \\
 Lesion volume (GG\_volume) & 0.21 & 0.18 \\
 Mean lesion HU value (GG\_MHU) & 0.058 & 0.22 \\
\bottomrule
\end{tabular}
\label{tab:corr}
\end{table*}


\section*{Discussion and Conclusion}

In this work, a novel way of predicting patient outcomes is proposed, namely admission to ICU and death, in COVID-19
patients by leveraging deep neural networks and ML classifiers. This work shows how deep learning can be leveraged to automatically extract quantitative metrics from lung CT scans and analyzes their impact on patient outcomes.

Two patient outcomes are predicted by considering 11 different multi-modal features - made up of 2 EHR-features and 9 image-features. While the EHR-features are directly read from the database, for extracting the image-features, three different DL models are re-trained / fine-tuned. Using the DL models various metrics related to lung segmentation (WL, NL, MCT, NLperc), muscle-fat segmentation (Muscle, Fat, MF ratio) and ground glass - consolidation segmentation (GG\_volume, GG\_frac, GG\_MHU) are extracted. To establish the validity of each of the trained segmentation models, a quantitative and qualitative evaluation of the model outputs is performed on respective test sets. The quantitative evaluation is performed using standard measures like Dice similarity and RVE, while the outputs are visually analysed by a Biomedical Image Analyst.

Finally, using EHR-features and image-features as inputs to various machine learning (ML) classifiers, the outcomes of COVID-19 patients are predicted. Specifically, logistic regression, SVM, random forest, ada-boost and XGBoost  are used as classifiers to model the outcomes. Using an internally curated dataset consisting of lung CT scans, the performance of ML classifiers is evaluated using AUC score. It is observed that  SVM and Random Forest are the best classifiesrs for predicting ICU admission which show a superior mean AUC of 0.77 compared to other classifiers. On the other hand for predicting death, Logistic Regression is the best classifier with a mean AUC of 0.73. 

While our reported performance metrics, with an AUC of 0.73 - 0.77 is less than a previous work which achieved an AUC of 0.85 \cite{zhang2020clinically} which utilized CT scans for predicting critical illness and an AUC of 0.91 using additional metadata, the efficacy of the model cannot be evaluated solely on the basis of AUC due to three main factors. Firstly, the difference in the definition of patient outcomes in different works. Secondly, the difference in distribution of patient characteristics (for instance, age). Thirdly, the patient admission criteria is different across countries and hence, a direct comparison is not ideal \cite{lassau2021integrating}. 

Besides this, using Gini importance, the relative importance of all the input features is shown. The relative importance of various features provides some critical insights into the features causing death / ICU admission. It is observed that age is by far the most important factor in predicting the death of a patient, followed by GG\_volume and GG\_frac. While there is no such clear determining factor for ICU prediction, a combination of factors related to the invasiveness of COVID lesions (GG\_MHU, GG\_frac), age and MCT predominantly determine the ICU admission. It is also observed that the fraction of the lung containing COVID lesions (GG\_frac) is an important feature which determines both the patient outcomes. 

Furthermore, the calculation of correlation between the features and the outcomes shows that increase in the value of features related to abnormalities like GG\_frac, GG\_volume and GG\_MHU increases the risk of death and ICU admission. Moreover, these features also have a high correlation with the outcomes. It was observed that males are at a higher risk of death and ICU admission and in addition, while older people are at higher risk of death, there is no strong correlation between age and ICU admission. Finally, on expected lines, features such as NL, NLperc, muscle and MF ratio are inversely proportional to the probability of death and ICU admission. Having a knowledge of the determining factors and the strength of the correlation is of paramount importance to determine patients at high risk of ICU admission and/or death.

One of the limitations of this work is the size of the dataset that was used to predict the patient outcomes. While the number of patients is objectively less, it is to be noted that it is difficult to obtain anonymised patient information containing both the EHR data and the CT scans. In the future, more patients will be added to the study as and when new data becomes available. and used for outcomes prediction  While modelling robust ML classifiers that can be deployed instantly to real world scenarios demand more data, the aim of this work is to show that it is beneficial to leverage the potential of medical imaging by coupling it with the traditional EHR data to build robust classifiers. We hope that such hybrid models draw the attention of ML researchers in the future. Moreover, only CT scans are currently used to extract image-related features for outcomes prediction. In the future, adding Chest X-rays as another source of image-related features will be explored. Finally, as a future research,  using the quantification methods for muscle and fat, the link between the progression of MF ratio and patient outcomes/disease progression will be explored. 

In conclusion, a framework is developed predicting the outcomes in COVID-19 patient outcomes using features read from EHR data and those extracted from CT scans using various DL models. The DL models used in extracting the image features are validated by performing quantitative and qualitative analysis. Next, the EHR-features and the image-features are used for training ML classifiers to predict patient outcomes. Finally, the performance of ML classifiers is evaluated using ROC curves to show the effectiveness of the proposed methodology.

\bibliography{sample}

\section*{Author contributions statement}
SVN and PC carried out the patient outcomes analysis experiments. CHL and PY developed the lung segmentation methodology and extracted the metrics for lung segmentation. AS developed the muscle fat segmentation and extracted the metrics for muscle and fat. SVN extracted the metrics for ground glass and consolidation. KT provided image  annotations for subsets of the data. MV and PS provided clinical direction and clinical requirements. BI designed the overall study and steps. MV, PS, BI, SVN analysed the results of the experiments. SVN wrote the paper with inputs from all the authors.

\section*{Acknowledgements}
This work uses data provided by patients collected by the National Health Service (NHS) as part of their care and support. We believe using the patient data is vital to improve health and care for everyone and would, thus, like to thank all those involved for their contribution. The data were extracted, anonymised, and supplied by the Trust in accordance with internal information governance review, NHS Trust information governance approval, and the General Data Protection Regulation (GDPR) procedures outlined under the Strategic Research Agreement (SRA) and relative Data Processing Agreements (DPAs) signed by the Trust and Sensyne Health plc. Special thanks to Bavithra Vijayakumar, Stephen Hughes, Mayowa Omosebi and Lionel Tarassenko.

\section*{Additional information}
 \textbf{Competing interests}:SVN, PC, CHL, PY, AS, KT and BI are employees of Sensyne Health that develops machine learning approaches for healthcare.
\end{document}